\documentclass[a4paper]{article}
\pdfoutput=1
\usepackage{icrc2013}
\usepackage[english]{babel}

\title{Multiwavelength study of the region around the ANTARES neutrino excess}

\shorttitle{F. Sch\"ussler for the H.E.S.S. Collaboration: ANTARES neutrino excess}

\authors{
Fabian Sch\"ussler$^{1}$,
P. Brun$^{1}$, R.C.G Chaves$^{1}$, J.-F. Glicenstein$^{1}$, K. Kosack$^{1}$, E. Moulin$^{1}$, B. Peyaud$^{1}$, D. Wouters$^{1}$ for the H.E.S.S. collaboration.\\
T. Stolarczyk$^{1}$, B. Vallage$^{1}$
}

\afiliations{
$^1$ Commissariat \`a l'\'energie atomique et aux \'energies alternatives / Institut de recherche sur les lois fondamentales de l'Univers \\
}

\email{fabian.schussler@cea.fr}

\abstract{The ANTARES collaboration reported the results of a search for point-like neutrino sources using data taken in the period 2007-2010. An unbinned maximum likelihood based all-sky search yielded a cluster of 9 (5) events within a cone of 3 (1) degrees around (R.A., Dec) = (-46.5deg, -65.0deg). The trial factor corrected p-value of $2.6\%$ (2.2 sigma) is not significant enough to claim the observation of an astrophysical point source. However, it currently constitutes the most significant localized neutrino excess observed by ANTARES. Here we present a multi-wavelength analysis including optical to X-ray archival data and a dedicated analysis of gamma-ray data from Fermi-LAT. In order to cover the TeV domain, dedicated observations with the H.E.S.S. telescope array were carried out. We present these data and discuss implications of the results in terms of signatures for a cosmic-ray acceleration site.}


\begin{document}
\maketitle

\section{Introduction}
The key question to resolve the long standing mystery of the origin of cosmic rays is to locate the sources and study the acceleration mechanisms able to produce fundamental particles with energies exceeding those of man-made accelerators by several orders of magnitude. Over the last years it has become more and more obvious that multiple messengers will be needed to achieve this task. Particle physics processes like the production and subsequent decay of pions in interactions of high energy particles predict, that the acceleration sites of high energy cosmic rays are also sources of high energy gamma rays and neutrinos.  Whereas more than 120 very high energy gamma-ray sources have been detected by ground based Cherenkov telescope arrays like H.E.S.S., MAGIC and VERITAS, no astrophysical neutrino source could be identified with high significance by the large scale neutrino telescopes like IceCube and ANTARES so far. Here we exploit the particle physics link between neutrinos and other messengers to study the region around the currently most significant high energy neutrino excess using data ranging from radio astronomy to very high energy gamma rays.\\

\subsection{The ANTARES neutrino excess}\label{AntaresExcess}
\begin{figure*}[!t]
   \centerline{\includegraphics[width=3.4in]{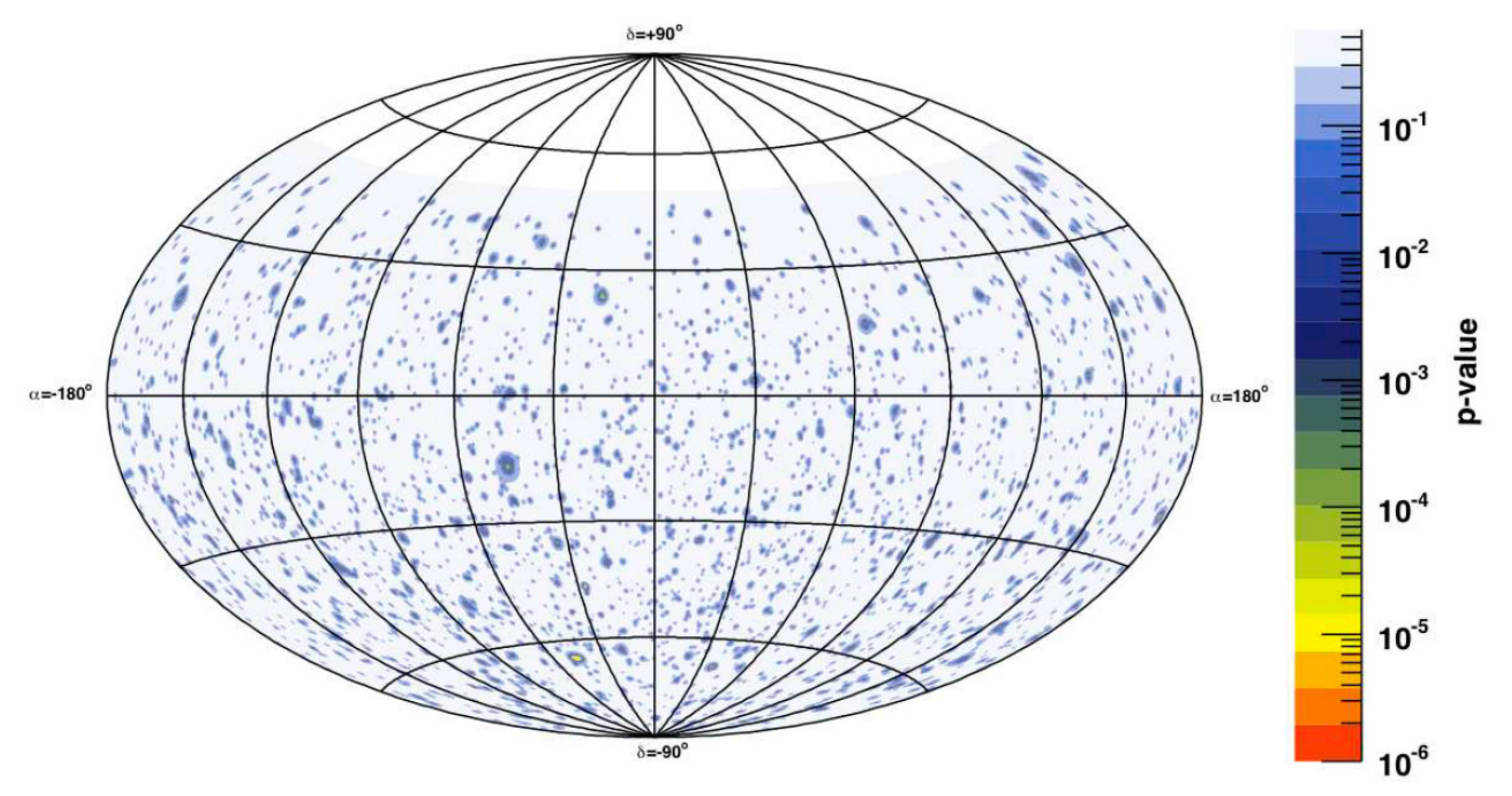}
              \hfill
              \includegraphics[width=3.4in]{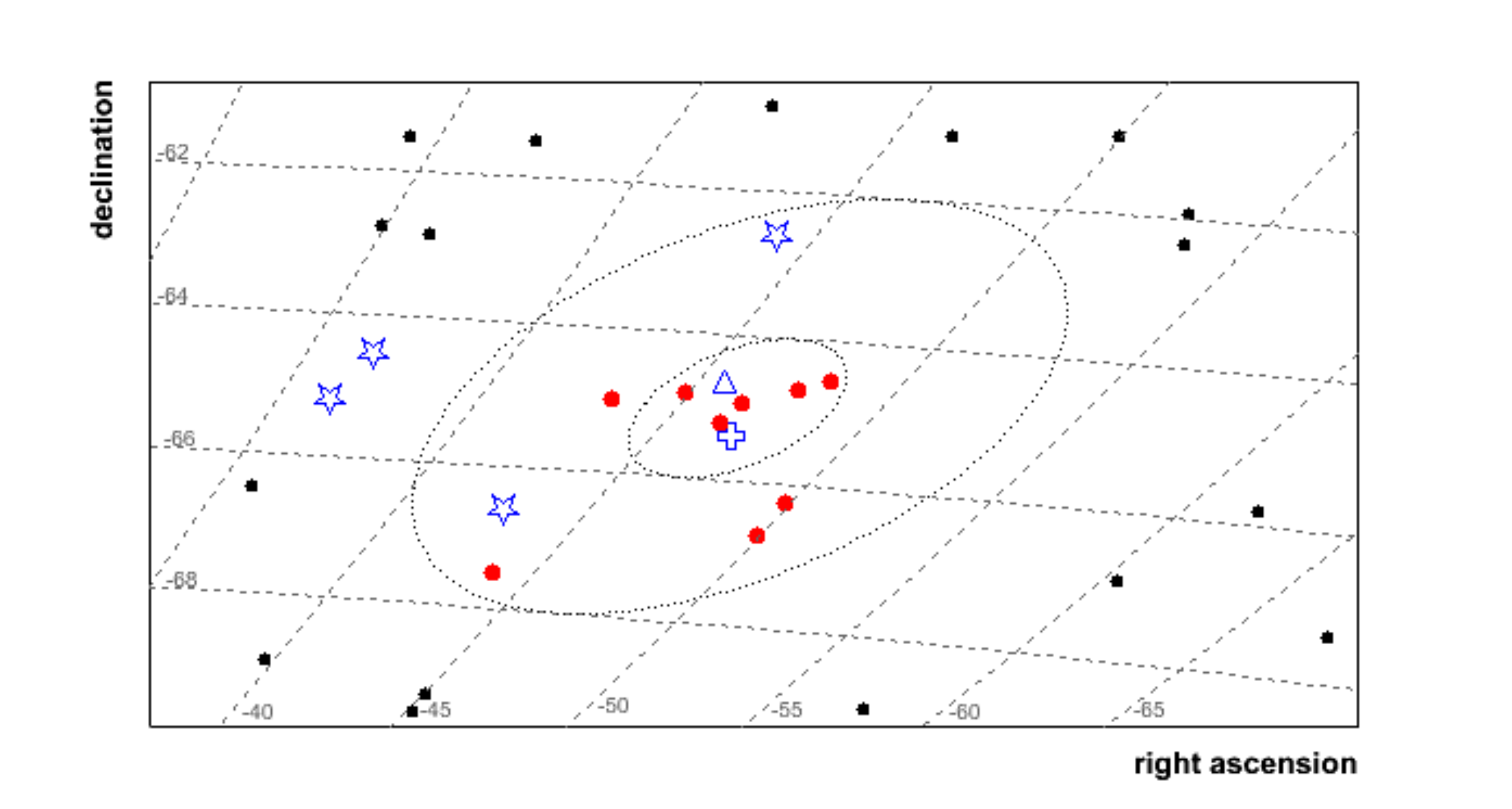} 
             }
   \caption{ANTARES neutrino excess. Left plot: Skymap in equatorial coordinates showing the p-values obtained in the all-sky search for point-like clusters. The penalty factors accounting for the number of trials are not yet considered in this representation~\cite{Antares_PS}. Right plot: Zoom in the location  of the most signal like cluster found in the full-sky search (equatorial coordinates). As indicated by the red circles, 5 (9) neutrino candidates are found within 1 (3) degrees of the cluster center as expected for a point-like source. Gamma-ray sources given in the 2nd Fermi-LAT catalogue are indicated by blue stars, PKS 2047-655 is denoted by the blue cross and AC 103 by the blue triangle.}
   \label{fig:AntaresExcess}
 \end{figure*}

The ANTARES neutrino telescope~\cite{Antares_DetectorPaper} started data taking in 2007 and is operating in its full configuration since 2008. The geometry and size of the detector makes it sensitive to neutrinos in the energy range from about 1 TeV to several 100 TeV. The ANTARES collaboration recently reported on a search for point like accumulations exceeding the isotropic atmospheric background~\cite{Antares_PS}. The analysis was based on optimal event selection criteria that had been developed using Monte Carlo simulations before unblinding the relevant data. The median uncertainty on the reconstructed neutrino direction assuming an $E^{-2}$ neutrino energy spectrum is $0.5 \pm 0.1~\mathrm{deg}$ and mis-reconstructed atmospheric muons only contribute at the level of $14~\%$ to the final data sample. Applied to the data recorded by ANTARES between beginning of 2007 and end of 2010 (corresponding to a total lifetime of 813 days) 3058 events pass the optimized selection criteria. An unbinned maximum likelihood method searching for point like high energy sources allowed to construct the significance map for the sky visible by ANTARES shown in Fig.~\ref{fig:AntaresExcess}.

The most significant cluster of events was found at $\mathrm{RA}=313.5^\circ, \mathrm{Dec}= -65.0^\circ$. As can be seen in the right plot of Fig.~\ref{fig:AntaresExcess}, 5 (9) events have been found within a cone of 1 (3) degree radius. For this cluster the likelihood fit assigns 5.1 signal events, compatible with the signature of a point-like high energy source. Pseudo-experiments taking into account systematic uncertainties of the angular resolution and the acceptance of the detector were used to determine the trial factor corrected p-value of $2.6\%$ (i.e. $2.2~\sigma$ using the two-sided convention). Given this significance, the ANTARES collaboration does consider the observed excess as compatible with a background fluctuation. 

However, taken at face value, i.e. considering a signal of 5 neutrino events as provided by the maximum likelihood fit, an approximate estimation of the corresponding neutrino flux can be given. Based on the effective area that has been derived from Monte Carlo simulations~\cite{Antares_PS}, and assuming an $E^{-2}$ energy spectrum, 5 events at a declination of $\delta=-65.0^\circ$ correspond roughly to a neutrino flux arriving at Earth of $\Phi_\nu \approx 5.5 \times 10^{-11}~\mathrm{TeV}^{-1}\mathrm{cm}^{-2}\mathrm{s}^{-1}$ with $80~\%$ of the events being in the energy range from 4 to 700 TeV.\\

\section{Archival multi-wavelength data}
Archival data of various wavelengths ranging from radio to UV and X-rays have been scanned in the search for a counterpart of the neutrino excess.~\cite{Simbad}. The region of interest contains the AGN PKS 2047-655 at a distance of $0.54~\mathrm{deg}$ from the excess center and the galaxy cluster AC 103 (ABELL S0910) at a distance of $0.87~\mathrm{deg}$. PKS 2047-655 has first been reported within the Parkes radio survey~\cite{ParkesSurvey} and is located at a redshift of $z=2.3$. It is a flat spectrum radio quasar without any particular difference to the common population of these objects. It is depicted by the blue cross in the right plot of Fig.~\ref{fig:AntaresExcess}. The galaxy cluster AC 103 has first been reported in~\cite{AC103}. It is a comparably small cluster located at redshift $z=0.31$ without any particularly striking feature. It is indicated by the blue triangle in the right plot of Fig.~\ref{fig:AntaresExcess}. Further searches in archival data did not yield hints about sources suspected to be able to accelerate particles to high energies.\\

\section{High energy gamma-ray data from Fermi-LAT}
Using two years of high energy gamma-ray data from the Fermi-LAT satellite, the Fermi collaboration compiled the 2FGL gamma-ray catalogue~\cite{Fermi_2FGL}. Sources found in this all-sky sample are indicated by the blue stars in the right plot of Fig.~\ref{fig:AntaresExcess}. No 2FGL source is located within 2 degrees from the center of the excess and no clear correlation with the location of the Antares excess can be established. 

\begin{figure*}[!t]
   \centerline{
   \hfill
   	\includegraphics[width=0.37\linewidth]{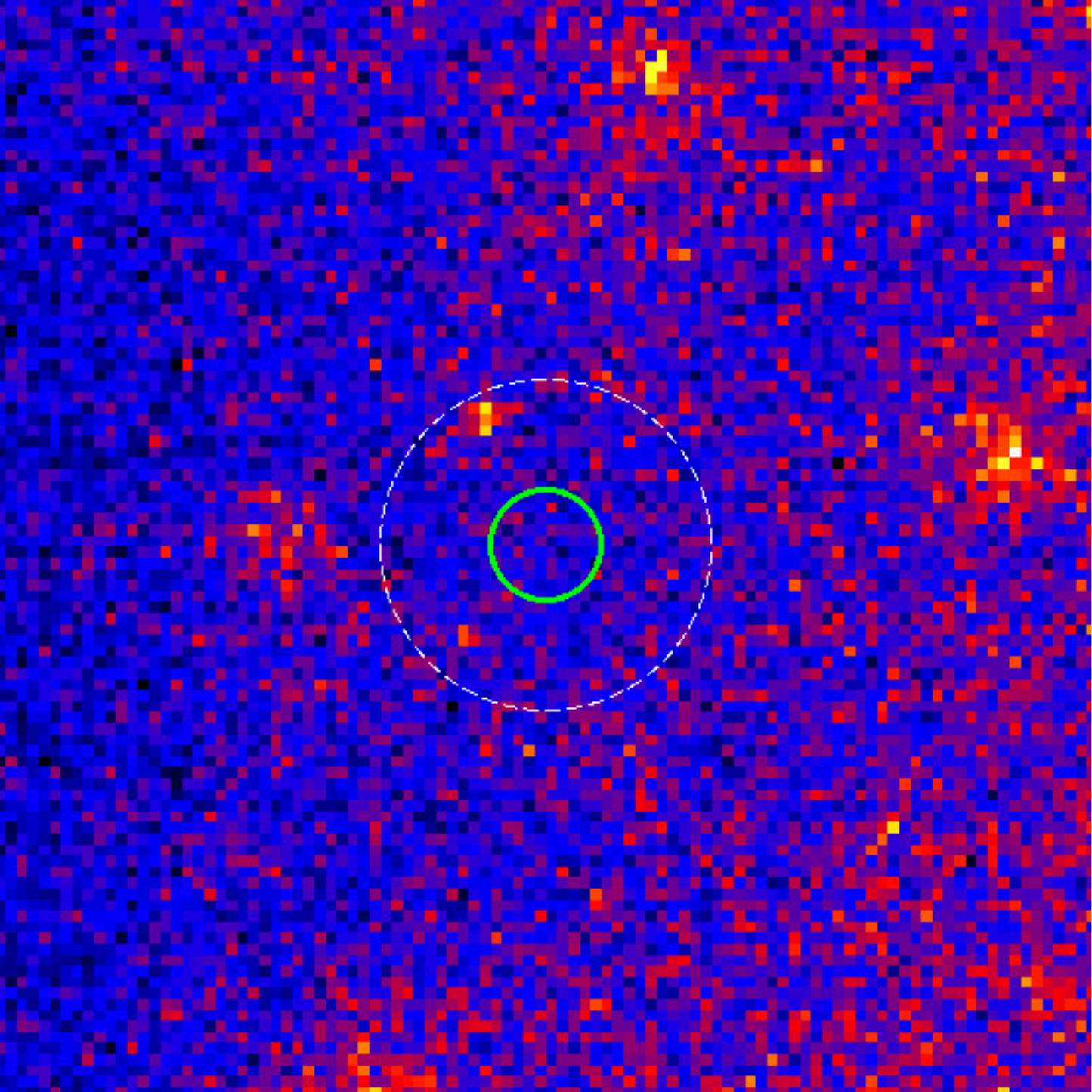}
              \hspace{0.1\linewidth}
              \includegraphics[width=0.37\linewidth]{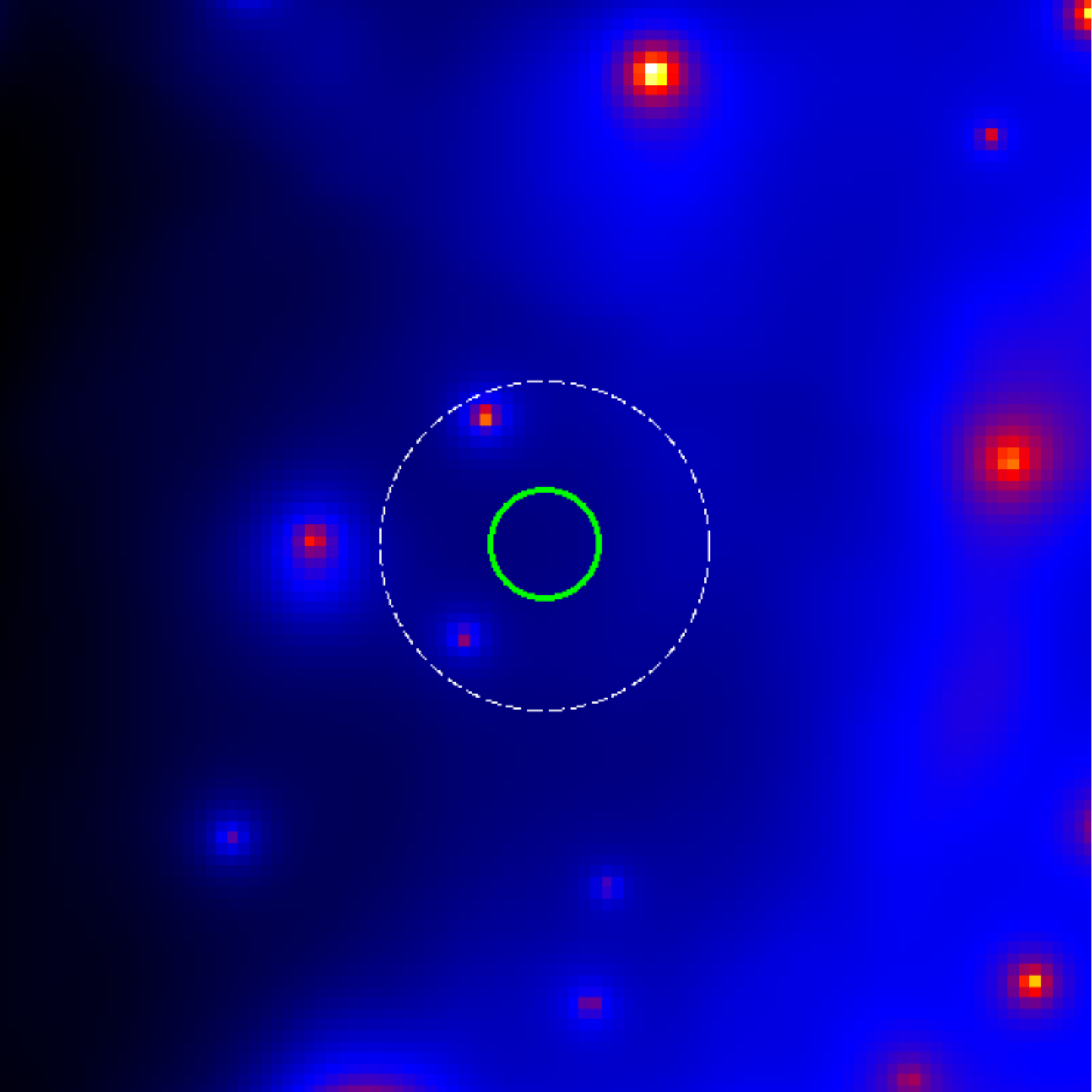} 
              \hfill
             }
   \caption{Region around the ANTAREs excess as seen by the Fermi-LAT gamma-ray observatory. The solid (dashed) circle denote the 1 (3) degree radius region of interest. Left plot: Count map of gamma-rays detected by Fermi-LAT in equatorial coordinates. Events fullfill the standard Fermi-LAT data selection requirements and cover the full energy range starting at $100~\mathrm{MeV}$. Right plot: Map showing the model of gamma-ray sources fitted to the Fermi-LAT data.}
\label{fig:Fermi}
 \end{figure*}

Doubling the dataset with respect to the one used for the 2FGL, we performed a dedicated analysis of 4 years of data from the Fermi-LAT instrument in order to search for potential gamma-ray emission at flux levels lower than the sources included in the 2FGL catalogue. The analyzed data was taken between 2008-08-04 and 2012-08-31. We selected all photon candidates (evclass=2) above $E=100~\mathrm{MeV}$ fulfilling the basic quality criteria proposed by the Fermi collaboration (quality=1; LAT config=1; rock angle$<52~\mathrm{deg}$; zenith$<100~\mathrm{deg}$) within a $10\times 10~\mathrm{deg}$ region of interest centered at $\mathrm{RA}=313.5^\circ, \mathrm{Dec}= -65.0^\circ$. The count map of the selected gamma-ray events is shown in the left plot of Fig.~\ref{fig:Fermi}.

The gamma-ray emission observed in the Fermi-LAT data has been modelled using the information given in the 2FGL catalogue. All parameters for sources further away than $5~\mathrm{deg}$ from the center of the excess have been fixed to the 2FGL values. An exception is 2FGL J1940.8 which is located at the eastern boundary of the region of interest (ROI) and has a low energy spacial extension that required a re-fitting of its parameters to arrive at a good description of the data. The extragalactic diffuse model (version p7v6) and the model of the Large Magellan Cloud located about $20~\mathrm{deg}$ away from the ROI have been taken into account. An additional point-like source with a power-law energy spectrum has been added to the description before fitting the model parameters to the count map of the selected events. The fit did not find significant emission from the additional putative source related to the ANTARES excess. The derived parameters for the sources (except 2FGL J1940.8) already found in the 2FGL are fully compatible with the parameters given therein.\\

\section{Very high energy gamma-ray data from H.E.S.S.}
\begin{figure}[!h]
  \centering
  \includegraphics[width=\linewidth]{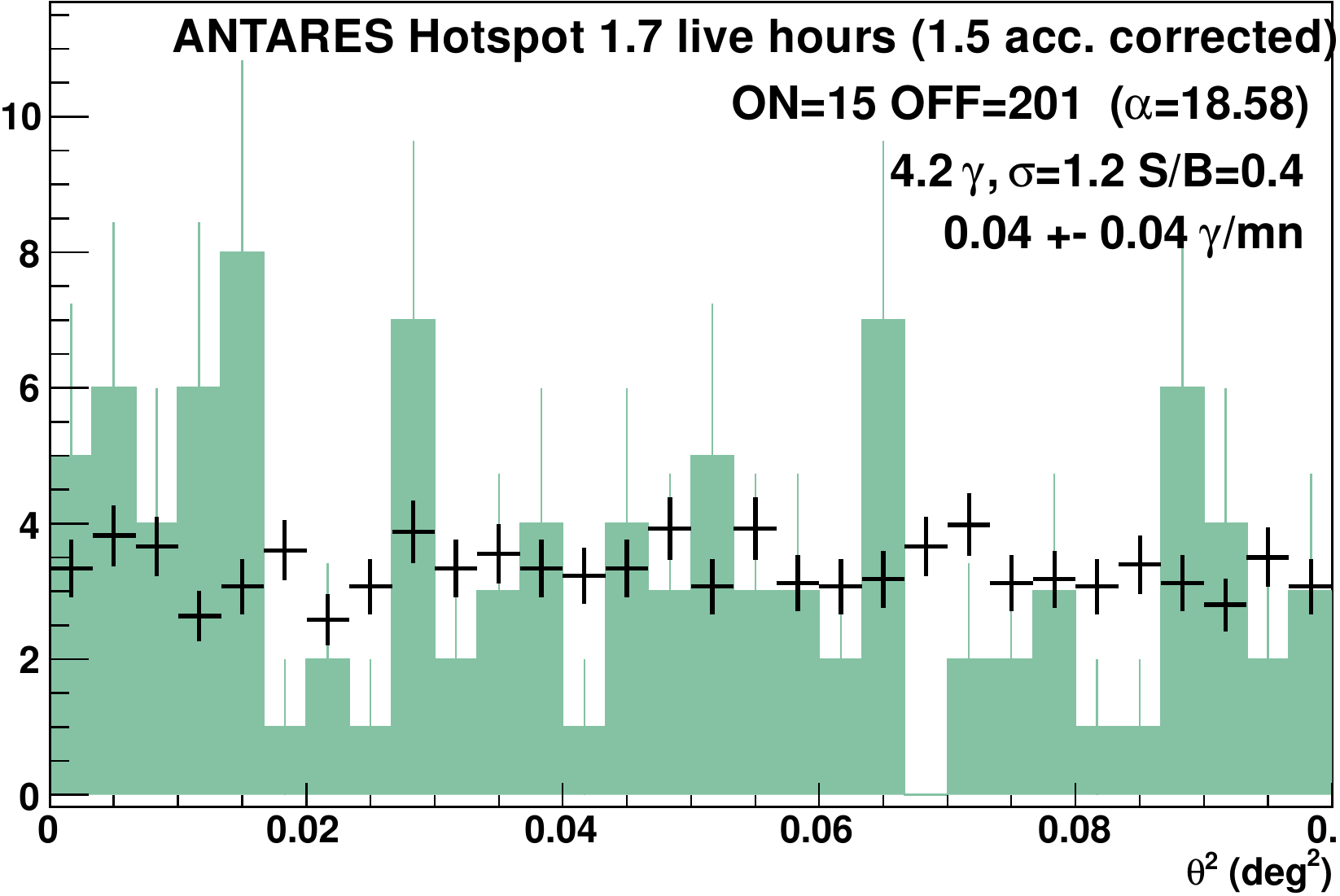}
  \caption{Distribution of gamma-ray candidate events as a function of their distance to the center of the ANTARES excess. The green histogram shows events from the "source" region and the black markers denote the background expectation.}
\label{fig:Theta2}	
 \end{figure}
 
The region around the neutrino excess has been observed by the H.E.S.S. high energy gamma-ray telescope system in its original configuration of four telescopes. In this setup H.E.S.S. is sensitive to detect cosmic and gamma-rays in the 100 GeV to 100 TeV energy range and covers a field of view of $5^\circ$ in diameter. Data has been taken for almost $2~\mathrm{h}$ in November 2012 at a zenith angle of $45^\circ$ in so-called Wobble-mode, pointing at 4 different positions offset by $0.7~\mathrm{deg}$ from the center of the ANTARES excess. After correcting for acceptance effects the effective lifetime corresponds to $1.5~\mathrm{h}$.\\

\begin{figure*}[!t]
  \centering
  \includegraphics[width=\linewidth]{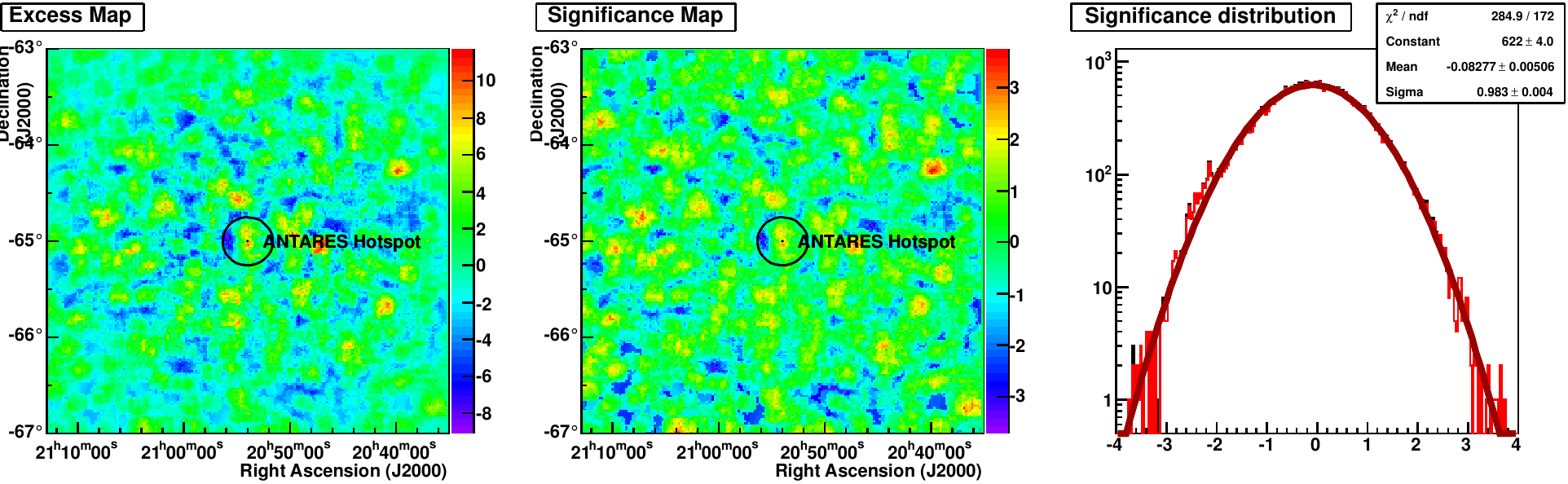}
  \caption{Left plot: Map of VHE gamma-ray events exceeding the background expectation. Middle plot: Point source significance map. Right plot: Distribution of point source significances in the field of view.}
\label{fig:HESS}	
 \end{figure*}

The data were analyzed using the Model Analysis~\cite{ParisAnalysis} with standard gamma-hadron separation and event selection cuts. The background has been determined using the "reflected background" method described in~\cite{RingBg}, a method that exploits the properties of the wobble data taking mode and yields very low systematic uncertainties related to the acceptance of the camera system. The distribution of the squared angular distance of gamma-ray candidates around the position of the ANTARES excess is shown in Fig.~\ref{fig:Theta2} for both the signal region (green histogram) and the background estimation (black markers). With only about 4 gamma-rays exceeding the background (corresponding to a significance of $1.2~\sigma$ following~\cite{LiMa}) within $\theta^2\leq 0.01~\mathrm{deg}$, no significantly enhanced gamma-ray flux towards the center of the ANTARES neutrino excess has been detected. The distribution of gamma-ray events exceeding the background is shown for the full ROI in the left plot of Fig.~\ref{fig:HESS}. The middle (right) plot of Fig.~\ref{fig:HESS} shows the map (distribution) of the Li \& Ma significances. They are fully compatible with the background expectation.\\

\subsection{Upper limits on the gamma-ray flux}
\begin{figure}[!h]
  \centering
  \includegraphics[width=0.98\linewidth]{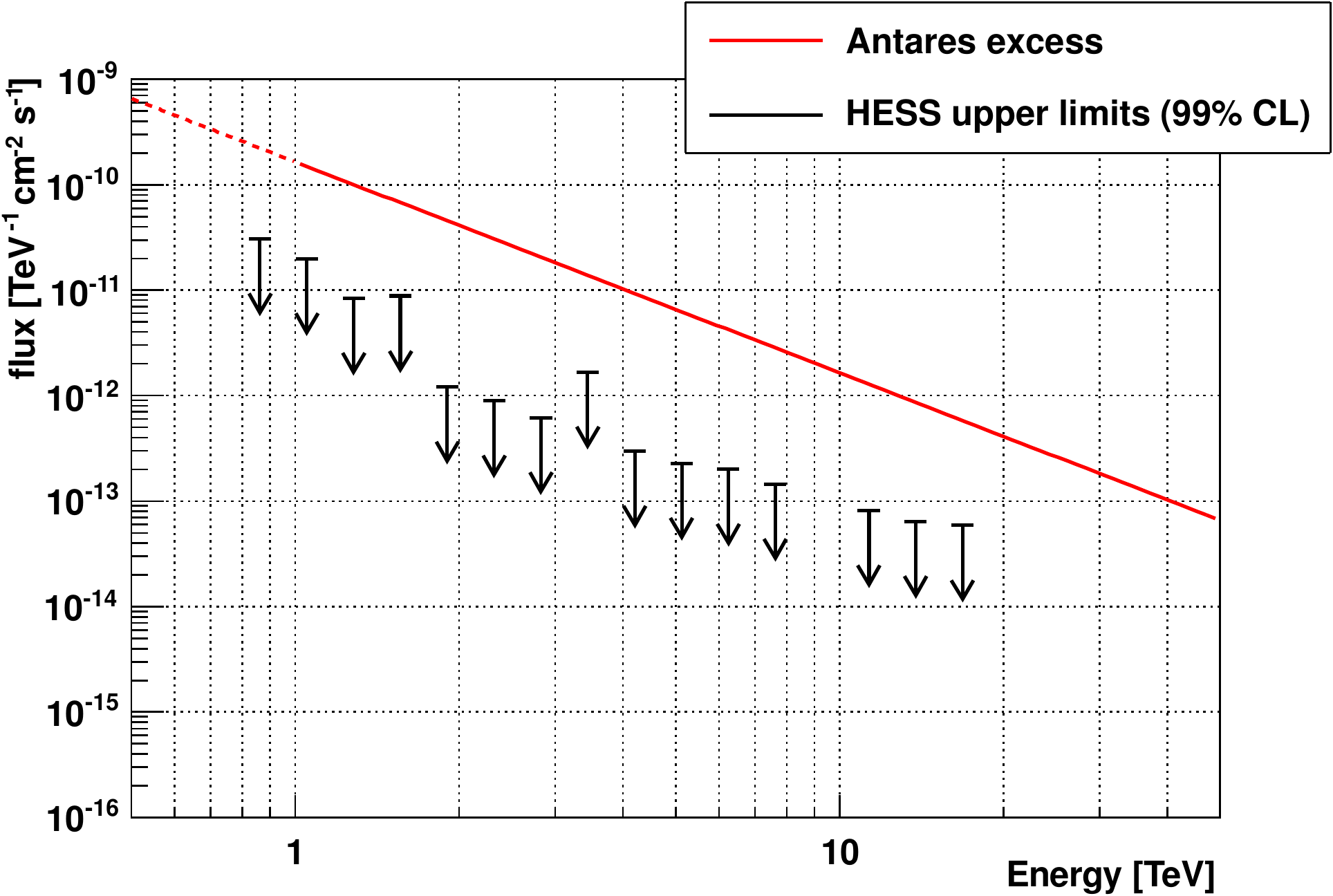}
  \caption{VHE gamma-ray flux limits $\Phi_\mathrm{UL}$ at 99~\% CL derived from the H.E.S.S. observations (black arrows) compared to predictions based on the ANTARES neutrino excess $\Phi_\gamma$ (red line). An $E^{-2}$ energy spectrum has been assumed.}
\label{fig:FluxLimits}	
 \end{figure}
 
Given the absence of a significant very high energy gamma-ray signal in the observed region we derived upper limits on the gamma-ray flux. The relatively high zenith angle of $45~\mathrm{deg}$ of the observations yields an energy threshold of around $800~\mathrm{GeV}$. The obtained flux limits $\Phi_\mathrm{UL}$ have been calculated assuming a generic $E^{-2}$ energy spectrum and following the method introduced by Feldman \& Cousins~\cite{FeldmanCousins}. The obtained $99~\%$ confidence level limits from the H.E.S.S. observations are shown as black arrows in Fig.~\ref{fig:FluxLimits}. The red line shows the expectation from the neutrino candidate events that has been derived by converting the neutrino flux of $\Phi_\nu \approx 5.5 \times 10^{-11}~\mathrm{TeV}^{-1}\mathrm{cm}^{-2}\mathrm{s}^{-1}$ (see Sec.~\ref{AntaresExcess}) into an associated flux of gamma rays. This conversion relies on Monte Carlo simulations of the hadronic interactions connecting neutrino and gamma ray fluxes via the decay of charge and neutral pions within or close to a generic hadronic accelerator. Following the assumptions and considerations given in~\cite{Kappes_NuGammaFlux}, the gamma ray flux produced by the source that produced the ANTARES neutrino signal can be estimated to about $\Phi_\gamma \approx 1.4 \times 10^{-10}~\mathrm{TeV}^{-1}\mathrm{cm}^{-2}\mathrm{s}^{-1}$ at 1 TeV.\\

\subsection{Lower limits on the source distance}
\begin{figure}[!h]
\vspace{-0.35cm}
  \centering
  \includegraphics[width=0.95\linewidth]{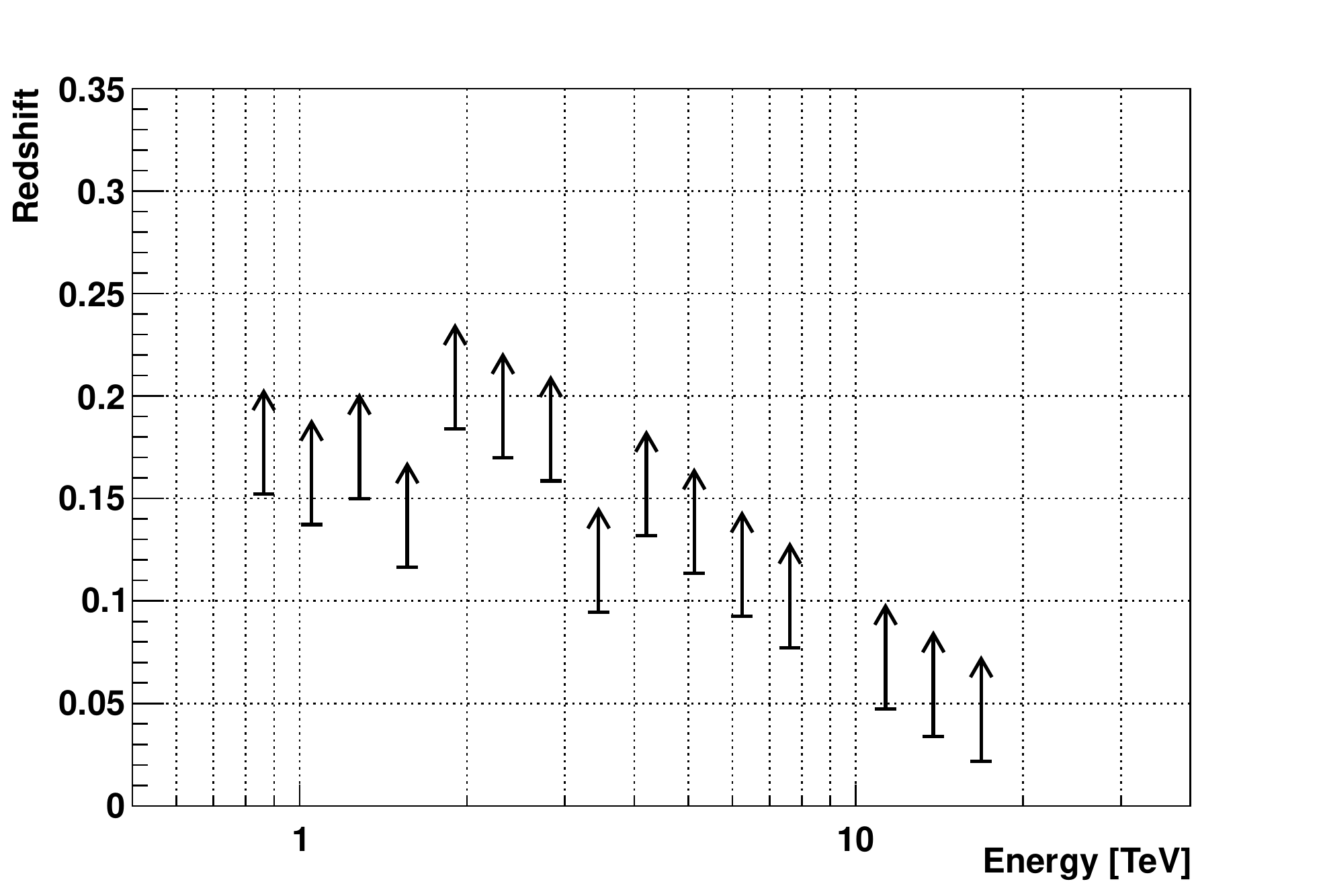}
  \caption{$99~\%~\mathrm{C.L.}$ lower limits on the distance of the potential neutrino and gamma-ray source derived by matching the ANTARES flux with the upper limits obtained with H.E.S.S.}
\label{fig:RedshiftLimits}	
 \end{figure}

High energy gamma-ray photons are absorbed by pair production on the extra-galactic background light (EBL). This process can be described by $\Phi_\mathrm{obs}=\Phi_\mathrm{source} \times e^{-\tau}$, where the optical depth $\tau$ is a function of the energy $E_\gamma$ and the redshift of the source $z_\mathrm{s}$. The photon density $n_z(\epsilon)$ as function of the photon energy $\epsilon$ is taken from the EBL model given in~\cite{EBL_Franceschini}, scaled with $k=1.27$ to match the H.E.S.S. measurements in~\cite{EBL_HESS}. $\tau$ can be written as:
$$
\tau(E_\gamma, z_\mathrm{s}) = \int_0^{z_\mathrm{s}} \mathrm{d}l(z) \int_{\epsilon_0}^{\infty} \mathrm{d}\epsilon \;\; \sigma_{\gamma\gamma}(E_\gamma(z+1),\epsilon) \times k \times n_z(\epsilon)
$$
For sufficiently distant sources, i.e. sufficiently large optical depths, the expected gamma-ray flux $\Phi_\gamma$ will get absorbed and will therefore become compatible with the upper limits $\Phi_\mathrm{UL}$ derived from the H.E.S.S. measurements. We exploit this possibility to derived lower limits on the distance to the putative neutrino and gamma-ray sources by solving the following equation for the redshift $z_\mathrm{lim}$ for all energy bins $i$:
$$
\tau(E_i, z_\mathrm{lim}) = -\ln\frac{\Phi_\mathrm{UL,i}}{\Phi_\gamma}
$$
The resulting $99~\%~\mathrm{C.L.}$ limits are shown in Fig.~\ref{fig:RedshiftLimits}.

\vspace{-0.5cm}
\section{Summary and conclusion}
We studied the region around the ANTARES neutrino excess using data ranging from radio observations to high energy gamma-rays detected by Fermi-LAT and H.E.S.S. No astrophysical source capable of producing the neutrino events has been detected in these additional messengers. The ANTARES excess seems therefore likely to be due to a background fluctuation.\\


{\footnotesize{{\bf Acknowledgment:}{Please see standard acknowledgement in H.E.S.S. papers, not reproduced here due to lack of space. We thank our colleagues of the ANTARES Collaboration for valuable discussions.}}}\\

\end{document}